\documentclass[%
 reprint,
 amsmath,amssymb,
 aps,
pra,
]{revtex4-2}

\usepackage{graphicx}
\usepackage{dcolumn}
\usepackage{bm}
\usepackage{hyperref}

\begin{document}

\preprint{APS/123-QED}

\title{The Application of Tailored Fields for Studying Chirality and Symmetry}

\author{Dino Habibovi\' c$^{1}$}

\email{e-mail: dhfizika1@gmail.com} 

\author{Kathryn R. Hamilton$^{2}$} 

\email{e-mail: kathryn.r.hamilton@ucdenver.edu}
\author{Ofer Neufeld$^{3}$}

\email{e-mail: oneufeld@schmidtsciencefellows.org}
\author{Laura Rego$^{4,5}$}
\email{e-mail: laura.rego@inv.uam.es}

\affiliation{$^1$University of Sarajevo, Faculty of Science, 71000 Sarajevo, Bosnia and Herzegovina}
\affiliation{$^2$Department of Physics, University of Colorado Denver, Denver, Colorado 80204, USA}
\affiliation{$^3$Max Planck Institute for the Structure and Dynamics of Matter and Center for Free-Electron Laser Science, Hamburg 22761, Germany}
\affiliation{$^4$Instituto Madrileño de Estudios Avanzados en Nanociencia (IMDEA Nano), Cantoblanco 28049, Madrid, Spain}
\affiliation{$^5$Departamento de Química, Universidad Autónoma de Madrid, 28049, Madrid, Spain}

\begin{abstract}
Ultrashort laser pulses pose unique tools to trigger and probe the fastest charge dynamics in matter, allowing the investigation of fundamental physical phenomena with unprecedented resolution in space, time, and energy. One of the most fascinating opportunities that ultrashort pulses offer is the possibility of modulating and investigating symmetries by tailoring the properties of the laser beam in the spatial and polarization domains, effectively controlling symmetry breaking on multiple levels. In particular, this allows probing chiral matter and ultrafast chiral dynamics. In recent years, the development of highly sensitive approaches for studying chirality has been a hot topic in physics and chemistry that has developed largely separately from the field of tailored light. This perspective discusses the individual and joint evolution of these fields with an emphasis on how the fields have already cross-fertilized, opening new opportunities in science. We outline a future outlook of how the topics are expected to fully merge and mutually evolve, emphasizing outstanding open issues.

\end{abstract}

\flushbottom

\maketitle

\thispagestyle{empty}

\section*{Introduction}

During the past few decades, laser technology has substantially advanced, resulting in the ability to generate and finely control multiple degrees of freedom of coherent light. One particularly important development is the production of ultrashort laser pulses. To date, the shortest laser pulses with durations of a few tens of attoseconds \cite{Agostini2004,Sansone2006,Orfanos2019} are generated table-top via the process of high-order harmonic generation (HHG) (a discovery recently awarded the Nobel prize in Physics \cite{Nobel1}), or from x-ray free electron lasers (XFEL) \cite{Serkez2018}. Such pulses provide unprecedented resolution in space, time, and energy \cite{Popmintchev2012,Calegari2014,kraus2015,peng2019}, and effectively allow probing charge dynamics in their natural timescales. This provides insight into fundamental physical phenomena such as electron correlations \cite{Kraus2018,Shiner2011,Valmispild2024,Freudenstein2022,Azoury2019,Wahyutama2019}, photoionization time delays \cite{Maquet2014,Grundmann2020,Schultze2010,Kheifets2020}, and coupled electron-nuclear dynamics \cite{Neufeld2022,Baker2006,Saito2019,Feng2012}. They also pave the route to exploring phenomena that intersect with applications such as charge migration and transfer \cite{Calegari2014,kraus2015,Bruner2017,Nisoli2017,Worner2017}, photocurrent generation \cite{Higuchi2017,Schultze2013,Schiffrin2013,Boolakee2022}, and molecular chirality \cite{beaulieu2017,Rozen2019,Cireasa2015}.

Immense scientific effort has been dedicated to the control over a wide variety of characteristics of ultrashort laser pulses, culminating in the prominent field of tailored light \cite{Dunlop2017,Shen2023,Bliokh2023,Hughes2021}. Such control has evolved over the years from fine manipulation of the pulse carrier frequency, bandwidth, pulse duration and envelope \cite{Demtroder2002,Diels2006,Salah2007,Weiner2011}, to more intricate details such as the beam spatial profile or its angular momenta, which can take two forms: spin angular momentum \cite{Poynting1909} (polarization) and orbital angular momentum \cite{allen1992} (vorticity). In particular, by incorporating multiple carrier waves into a single coherent beam, one gains immediate control of light's symmetry (and asymmetry) properties \cite{neufeld2019}. This emerged initially in the form of bi-chromatic bi-circular fields \cite{milosevic2000, eichmann1995, Long1995}, which also allowed the generation of circularly-polarized x-rays through HHG \cite{fleischer2014, kfir2014, JimenezGalan2018, habibovic2020, habibovic2021}, and more recently allowed intricate waveforms for exploring molecular chirality \cite{ayuso2019,Neufeld2023}. At the ultimate regime of control, all of these tailored properties could be tuned simultaneously to generate waveforms with coupled degrees of freedom, i.e. where the symmetries and symmetry breaking of the beam derive from a non-trivial combination of the spatial, temporal, and polarization degrees of freedom \cite{Pisanty2019,lerner2023,Rego2019,Sederberg2020,Tzur2022}. In combination with the field of HHG, by driving non-linear response with such tailored beams, their physical properties can be imprinted onto generated attosecond beams and X-ray emission. This paves the way to exploiting the beams' unique symmetries and asymmetries for probing fundamental phenomena both directly and in down-the-line attosecond spectroscopies.

A separate field of research that has been actively perused in recent years is the field of chirality. Chirality is a universal property in nature, emerging on all length scales from galaxies \cite{Kondepudi2001} to snail shells \cite{Maderspacher2016}, and down to fundamental particles \cite{Griffiths2008}. 
In general, an object is chiral if and only if it can not be superimposed onto its mirror image, connecting the physics behind chirality to symmetries \cite{Bishop2012}. At the nanoscale, chirality manifests in the form of chiral molecules, whose two opposite mirror-reflected versions are denoted as enantiomers. Oppositely-handed enantiomers are generally indistinguishable, except for when interacting with another chiral object (e.g. in a chemical reaction with other chiral molecules, or when interacting with circularly-polarized light). 

Since most biological molecules are chiral, chirality is extremely relevant in biology and pharmaceutical science, where highly sensitive identification of chiral molecules is a current open problem. Chirality is also of interest in particle physics \cite{Erez2023,Quack2002}. 
Traditional optical methods to distinguish chirality are based on interactions with elliptically polarized light, relying on the interplay between electric-dipole and magnetic-dipole interactions, and thus producing weak signals \cite{Berova2013}. In recent years there has been a massive effort from the physics and chemistry communities to improve the sensitivity of chiral detection schemes using a variety of approaches. Among these approaches are the use of nonlinear light-matter interactions \cite{Simpson2004,Fischer2005}, utilizing X-ray wavelengths \cite{Zhang2017}, microwave wavelengths \cite{Patterson2013,Patterson2013b}, or even high laser intensities to drive chiral HHG and photoionization \cite{Cireasa2015,Bovering2001,janssen2014}. More recently, these efforts have begun to merge with the community of tailored light, e.g. utilizing helical dichroism from topological beams \cite{Forbes2018,Brullot2016,Rouxel2022,Ye2019}, employing polarization tailored light \cite{Neufeld2019_2,Rozen2019,Baykusheva2018}, and light with combinations of polarization and spatial phase tailoring\cite{ayuso2019,yachmenev2019,ayuso2021,ayuso2022,katsoulis2022,khokhlova2022,mayer2022}.

In this perspective, we review recent advances on these two fronts, emphasising how controlling symmetries in light fields allows for novel directions in the research of chirality. We also provide an outlook for future development in both fields, as well as new breakthroughs which could arise as a result of the synergistic study of tailored light and chirality.

\section*{Tailored laser fields}

The coherent nature of laser light allows for precise control over its properties. For example, one may combine different frequencies in a laser beam, harness the laser polarization, design beams with spatial structures, or even control the temporal shape of the pulse. These different kinds of laser fields can be encompassed under the term "tailored laser fields" and many light-matter interactions can be driven and controlled by them by utilizing their various degrees of freedom. In this section, we will review the recent developments in tailored fields in the context of ultrafast light-matter interactions. 

\begin{figure*}[ht!]
    \centering
    \includegraphics[width=0.8\textwidth]{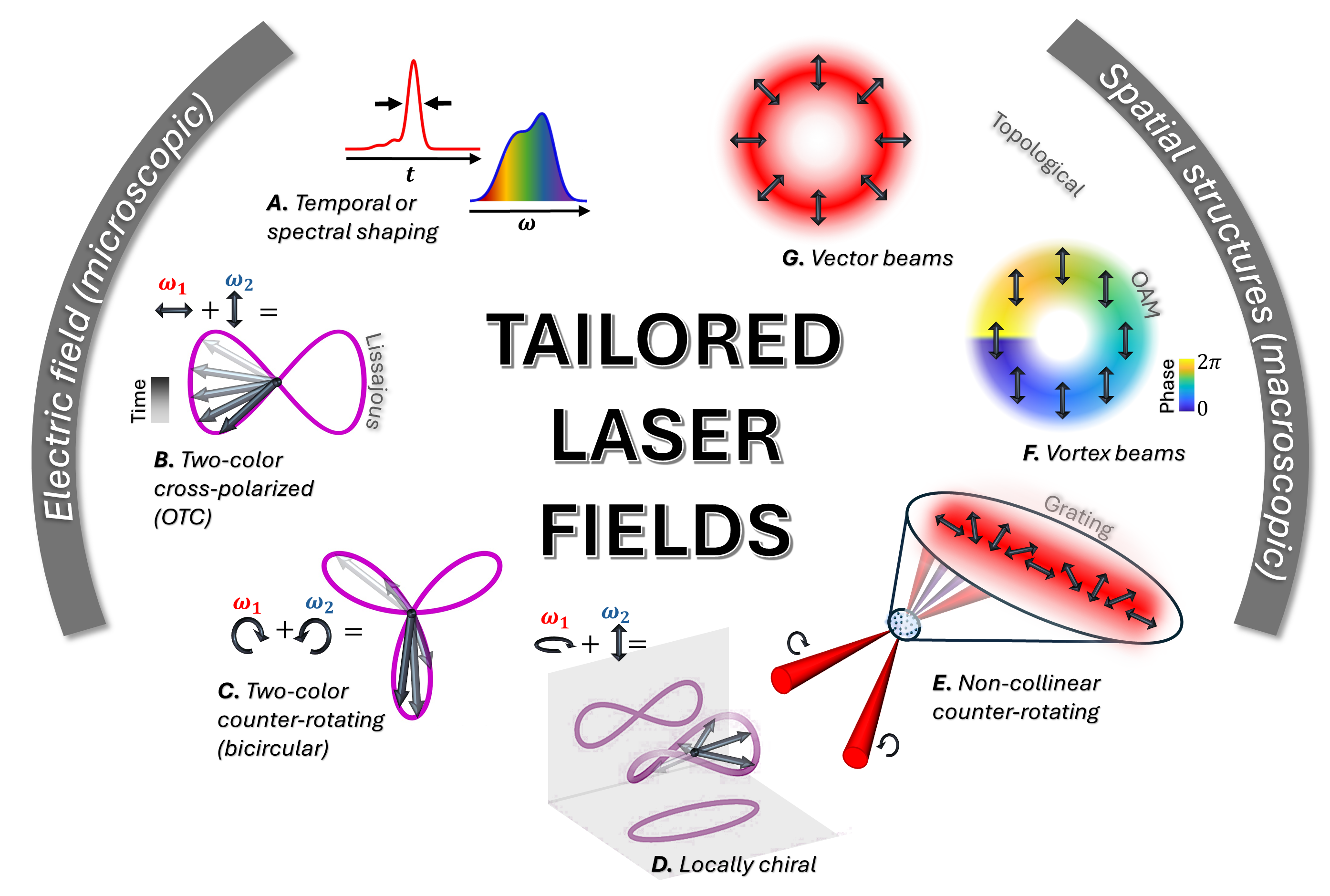}
    \caption{\textbf{Scheme of different types of tailored laser fields}. Generally, field tailoring may refer either to the control over the laser spectral, temporal, and polarization properties, or to the control over the spatial structure of the beam. Thus, it is possible to tailor microscopic properties, that can be defined at each point in space (left side of the figure), macroscopic properties arising from the laser's spatial variations (right side of the figure), or combinations of both. Conventional tailoring consists of the shaping of the temporal duration and spectral content of laser fields (panel \textbf{A}). Control over the laser polarization, beyond the generation of laser fields with different ellipticities, can be achieved by combining multiple frequency components, which results in an electric field drawing Lissajous figures in time. Two prominent examples are two linearly polarized components with orthogonal polarizations (panel \textbf{B}) and two circularly polarized components with opposite helicities (panel \textbf{C}). Three-dimensional local polarizations can also be created, such as in the locally-chiral light (panel \textbf{D}). Polarization can also be harnessed macroscopically, such as through the creation of spatial gratings, as shown in panel \textbf{E} for the typical set up of two non-collinear counter-rotating laser fields. Another important kind of macroscopic laser tailoring is the generation of light beams with topological properties, such as vortex beams, carrying an azimuthally-varying phase around a phase singularity (panel \textbf{F}) and vector beams, with azimuthally-rotating polarization around a polarization singularity (panel \textbf{G}). }
    \label{fig:tailored}
\end{figure*}

Laser tailoring is a very wide area of research and numerous kinds of light sources have been developed using a variety of tailoring techniques over the past decades. For illustrative purposes, Figure \ref{fig:tailored} shows a schematic summary of different categories of laser tailoring, but beams can also be generated using combinations of the described techniques. We can consider two main ways of taking control over the different attributes of laser light: (i) tailoring the electric field properties at the microscopic level (i.e. properties that can be defined at each infinitely small region of space, such as spectro-temporal properties or polarization, shown at the left side of Figure \ref{fig:tailored}), and (ii) tailoring the laser beam spatial shape, thus at the macroscopic level (i.e. properties that derive from the spatial variation of the field, such as topological light or polarization gratings, shown at the right side of Figure \ref{fig:tailored}). In addition, novel forms of light sources can be considered as a non-trivial combination of microscopic and macroscopic tailoring, such as polarization vortex beams that will be discussed below.

The first efforts in the control over laser field properties were mainly focused on harnessing its intensity (e.g. to generate nonlinear phenomena), as well as its spectral and temporal properties, creating laser light with tunable wavelengths and bandwidths (e.g. for spectroscopic applications), and aiming for the shortest durations to investigate ultrafast dynamics (Figure \ref{fig:tailored}A). This type of tailoring is also particularly relevant for nonlinear phenomena \cite{Hamilton2017}.

Other prominent types of tailored fields are beams obtained from coherent combinations of two or more carrier components with different frequencies and polarizations. Since these beams have multiple frequencies (typically a fundamental frequency and its second harmonic are employed due to convenience), their Lissajous curves can exhibit time-evolving polarizations with unique structures. That is, the polarization state of the beam can evolve in a non-trivial manner in time, generating and/or breaking specific symmetries of the electromagnetic field. Particularly important examples include orthogonally-polarized two-color fields (see Figure \ref{fig:tailored}B), which consist of two linearly polarized components with mutually orthogonal polarizations (allowing to generate mirror-symmetric and time-reversal invariant waveforms \cite{neufeld2019}) and bi-circular fields (see Figure \ref{fig:tailored}C), which consist of two circularly polarized components with different frequencies (allowing the generation of rotationally-symmetric waveforms\cite{fleischer2014,Reich2016,Kerbstadt2019,Eickhoff2021}). More generally, one can combine multiple carrier waves \cite{Li2019} or even non-commensurate waves \cite{Bandulet2010}, which present a playing ground for tuning symmetries in light's time-dependent polarization plane. 
Note that in poly-chromatic beams, besides frequency and polarization, the relative phases and intensities of the carrier-beam components are also available as control knobs that provide additional degrees of freedom often useful for ultrafast spectroscopy. 

Light's polarization can also be shaped in three dimensions by combining two frequencies with non-coplanar polarizations, e.g. one elliptically polarized field plus a linearly polarized field in the orthogonal direction (see Figure \ref{fig:tailored}D). This field is denoted as synthetic locally-chiral light \cite{ayuso2019, Neufeld2019_2}, as it has the peculiarity of drawing a chiral shape in time in every point in space, thus being locally-chiral. The interaction of such beams with chiral molecules will be described in next section. Importantly, such 3D polarizations can be generated using non-collinear schemes or tightly focused beams, where the resulting polarization state varies spatially and can interact with all of the sample's main axes as opposed to only in a plane.

This brings us to another family of tailored fields, those that carry macroscopic structures. Such tailored fields are often denoted as structured light, where a property of light -- such as intensity, phase or polarization -- varies spatially following a certain pattern. Among these are spatial gratings created in non-collinear schemes, resulting in periodic patterns in space (see Figure \ref{fig:tailored}E). For example, non-collinear counter-rotating driving beams with the same color create polarization patterns that allow for the generation of circularly polarized attosecond pulses \cite{Hickstein2015}. Macroscopic gratings can also generate optical Talbot carpets used for transient excitations \cite{Rouxel2021}, or induce waveforms with topological properties such as skyrmions\cite{Tsesses2018}.
 
A particularly interesting kind of spatially structured beams are those that carry orbital angular momentum (OAM) \cite{allen1992}. While the spin angular momentum (SAM) of light describes its polarization -- a microscopic property defined locally at each point of space -- the OAM of light is connected to the macroscopic shape of a paraxial light beam (paraxiality is required for separating these two degrees of freedom). OAM beams exhibit twisted phase front and doughnut-like intensity profiles around a phase point-singularity and are therefore often denoted as vortex beams (see Figure \ref{fig:tailored}F). 
Vortex beams offer exciting possibilities for particle manipulation \cite{Grier2003, Simpson1996}, information transfer \cite{Wang2012}, phase contrast \cite{Furhapter2005}, super-resolution microscopies \cite{Vicidomini2018}, and in quantum information \cite{Mair2001, Torres2011}.
The generation of OAM beams in the extreme-ultraviolet (EUV) or x-ray spectral regimes is motivated by the possibility of extending the current applications of vortex beams to the nanometric scale, especially in microscopy and spectroscopy \cite{Sakdinawat2007, vanVeenendaal2007, Picon2010, Baghdasaryan2019}. Vortex beams are typically produced in the optical and infrared regimes using spiral-phase plates, q-plates, holographic techniques or intracavity techniques. These methods however become inefficient for imprinting OAM to EUV or X-ray light, which are desirable for probing ultrafast phenomena. As alternatives, high-frequency beams carrying OAM have been generated via HHG \cite{Gariepy2014, Geneaux2016} or particle accelerators \cite{Woods2021}. In HHG, harmonic vortices exhibit topological charges that follow simple conservation rules \cite{Garcia2013b}, whereby the vortex nature of the beam is effectively imprinted onto the nonlinear optical emission. 
By combining temporal and spatial degrees of freedom together even more exotic characteristics can be induced, such as spatiotemporal optical vortices \cite{Jhajj2016, Chong2020} or time-varying OAM \cite{Rego2019}.

One noteworthy aspect of vortex beams is their connection to topology. A topological property of a geometric object is a quantity that is preserved under continuous deformations such as stretching, twisting, or bending. Light’s OAM is often denoted as a topological charge, whereas vortex beams are referred to as “topological light”. But, what exactly is topological about light? 
We can understand this connection by diving into its historical origins \cite{Dennis2001}. Young’s discovery of light’s wave nature at the beginning of the nineteenth century motivated the study of wave-related properties of light. Parallel to this, phase singularities started to be studied in different kinds of waves, such as in tides \cite{Whewell1833} (1833) or wavefunctions \cite{Dirac1931} (1930). However, the first connection between phase singularities in waves and topology appeared in 1974 in a seminal article by Nye and Berry \cite{Nye1974}, where they associated phase singularities in waves to wavefront dislocations, which can be studied in the same framework as crystal dislocations, opening the route towards attributing topological properties to light. From this point on, the term “topological charge” started to be used in vortex beams \cite{Tomita1986, Chiao1986, Coullet1989, Soskin1997}. This conserved charge is defined through the integral of the beam’s phase gradient in closed loop around the singularity: $\ell= \frac{1}{2 \pi} \oint_C \nabla \phi(\textbf{r}) dr$ \cite{Halperin1981}, where $\phi (\textbf{r})$ is the spatially-varying phase and $C$ denotes the loop.
Finally, in 1992, Allen et al \cite{allen1992} connected light’s phase singularities to OAM.

It is worth noting that phase singularities are not the only possible singularities in light beams --- polarization singularities are also very relevant, with vector beams the most well-known example \cite{Zhan2009}. Vector beams exhibit a nonzero Poincar\'e index, which is a topological index that describes the number of complete polarization rotations in a closed loop \cite{Gunther2021} (see Figure \ref{fig:tailored}G). Phase and polarization singularities can be further combined, such as in vector-vortex beams \cite{Zhang2017v} (also denoted as vectorial vortices \cite{Niv2006} or full Poincar\'e beams \cite{Beckley2010}), and polarization knots \cite{Leach2004, Dennis2010, Larocque2018}, where torus-knots can be described by their torus-knot angular momentum \cite{Pisanty2019}. Other 3D polarization structures include flying doughnuts \cite{Hellwarth1996, Papasimakis2018, Zdagkas2022, Jana2024}, where the longitudinal component of light creates a toroidal polarization shape, and polarization M\"obius strips \cite{Freund2010, Bauer2015}, where the polarization structure creates a single-side surface.

The connection between light and topology is interesting in the framework of light-matter interactions, as topological properties of light are connected to its symmetries, allowing us to define conserved quantities and design topologically protected (more robust under perturbations) configurations \cite{Vanderbilt2018}. However, it is important to also acknowledge the differences between light's topology and topology as it's often regarded and analyzed in the condensed-matter community, i.e. connecting to the Nobel prize in 2016 \cite{Nobel2}. The topology of materials leading to topological insulators, topological semimetals, protected surface states, and more, arises directly from the symmetry properties of the Material's Hamiltonian \cite{Vanderbilt2018}. Formally, material topology is a property defined in the ground state and under adiabatic evolution (with some recent extensions to out-of-equilibrium \cite{Rudner2020}). The topological protection of surface states appears as a result of an underlying symmetry (e.g. time reversal) and exists as long as that symmetry is respected, including by perturbations. In contrast, the conserved charges associated with light beams arise in the wave functions rather than from the fundamental Hamiltonian (i.e. in solutions to Maxwell's equations rather than from Maxwell's equations themselves). While the charges of a given beam are overall conserved, they are not offered exactly the same kind of protection as electronic wave functions in crystals. For instance, beams of high values of OAM might split up into many beams with lower OAM, or transfer their OAM to other entities through interactions \cite{Yao2011}. They also do not lead to another set of protected states as the concept of bulk-edge correspondence is not relevant for freely propagating beams. Lastly, we note that in topological light beams, the singularity is always removable in the sense that no physical property actually diverges at the singularity (because physical properties such as intensity or polarization exhibit a well-behaved limit, and the electromagnetic field is smooth and continuous everywhere). This differs from the situation in topological materials where there is a non-removable topological obstruction in k-space \cite{Thouless1984,Thonhauser2006}. At the same time, we should note that there are many analogue properties between topological photonics and topological condensed-matter, especially with using photonic crystals, which can also lead to new physical effects \cite{Ozawa2019,Rechtsman2013,Bandres2018,Lustig2019,Silveirinha2016,Maguid2017}.

From a practical standpoint, using tailored fields to drive laser-matter interactions may require special considerations compared to more standard light. First, from the theoretical point of view, complex polarization structures break typical symmetries associated with linearly polarized light, leading to more demanding calculations. On the other hand, it is important to note that modelling nonlinear light-matter interactions using spatially-structured light requires performing macroscopic calculations (at the very least in the plane transverse to the beam propagation), in contrast to standard cases where the single atom/molecule (point emitter) response is sufficient. This is because the spatially-varying response means that contributions from all local emitters need to be added up coherently and propagated towards the detector. Thus, the optical responses are strongly influenced by the spatial structure of the beam, adding sensitivity to micrometer-scale imperfections in the samples as well. Moreover, phase-matching effects might be relevant for very intense laser pulses or dense media along the propagation direction (requiring a macroscopic description of the interaction). In certain cases (such as chiral media discussed below) commonly employed approximations might break  \cite{Dreissigacker2014}. From the experimental perspective, this immediately means that one must consider the spatial three-dimensional geometry of the setup where it could also be crucial to adjust the position of the target with respect to the beam's waist. Often safely ignored effects such as the Gouy phase can all of a sudden play an important role \cite{Zhang2017v,Lopez2019}.

Finally, we may contemplate potential novel applications of tailored fields and the direction of this research area as a whole. In our opinion, polarization-tailored sources are especially interesting in the context of non-linear light-matter interactions and should prove highly useful for spectroscopy and coherent control in the coming years. For example, the two-color tailored fields employed to control the polarization of EUV or X-ray light \cite{eichmann1995, Long1995, fleischer2014, kfir2014, JimenezGalan2018,milosevic2000, habibovic2020,habibovic2021b,Habibovic2021c,habibovic2022,milosevic2023, Irfana2021} can also be employed to harness valley degrees of freedom \cite{Jimenez2020, Mrudul2021} and material topology \cite{Trevisan2022}. On the other hand, topological beams can serve as a spectroscopic tool to explore the target's anisotropy \cite{GarciaCabrera2024} and generate out-of-equilibrium states with novel properties \cite{Bhattacharya2022}, or even control optoelectronic responses \cite{Sederberg2020}. More generally, by taking advantage of the spatially-varying phase in OAM beams, one can hope to control different properties of the optical responses, as has been for instance already employed to control the polarization of the attosecond pulses \cite{Dorney2018} and their spectral content \cite{Rego2022}. 
Lastly, we highlight one extremely relevant field for applying such capabilities --- chirality. Given that chirality connects with the geometrical properties of objects, and tailored fields allow manipulating the geometry of light on multiple levels, the two should match extremely well. In the next section, we introduce chirality and the state-of-art examples of how it benefits from the use of a tailored laser field.

\section*{Bringing tailored fields to chirality}

\begin{figure*}[ht!]
    \centering
    \includegraphics[width=0.8\textwidth]{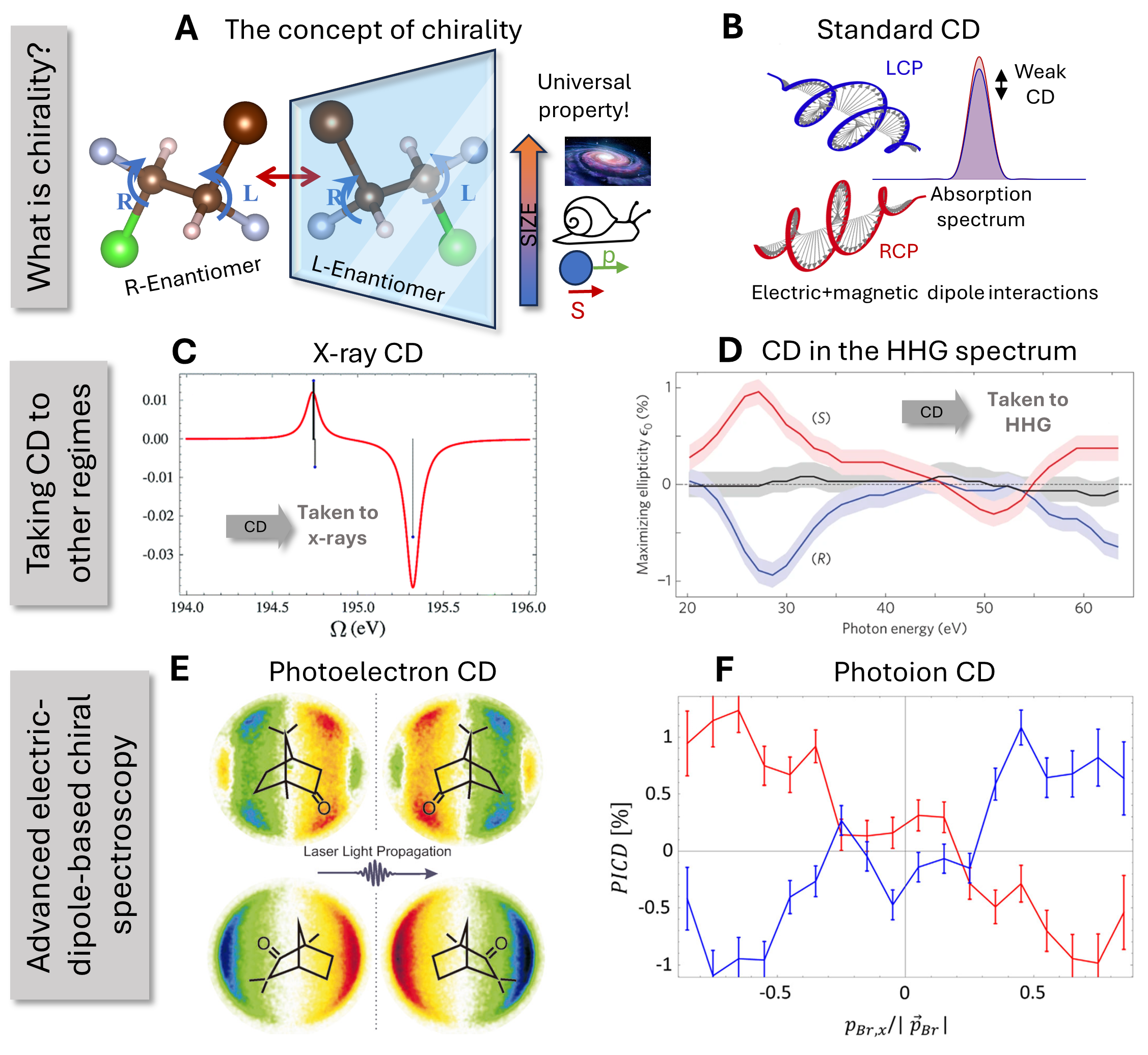}
    \caption{\textbf{The study of molecular chirality using standard laser fields}. Panel \textbf{A}: Scheme illustrating the concept of chirality, a universal property found at all length scales. Molecular enantiomers are the mirror-reflected versions of a chiral molecule. Panels  \textbf{B-D}: Traditional methods for the analysis of molecular chirality are based on CD, which measures the difference between the responses of the two enantiomers to elliptically-polarized light. Standard CD consists of measuring differences in absorption of light with right circular polarization (RCP) and left circular polarization (LCP), as illustrated in panel \textbf{B}. This concept can be extended to the x-ray regime, where the enantio-sensitivity is higher, as shown in panel \textbf{C}, figure adapted from \citenum{Zhang2017}. CD can also be found in the variation of the HHG yield with the ellipticity of the driving field, as presented in panel  \textbf{D} (adapted from \citenum{Cireasa2015}). Panels  \textbf{E-F}: Modern approaches to analyze molecular chirality based on electric-dipole interactions with standard circularly/elliptically polarized light. In PECD, the angular distribution of the photoelectron momentum shows an enantio-sensitive asymmetry in the direction of the laser field propagation, as shown in panel \textbf{E}, reprinted from \citenum{lux2012}. Enantio-sensitive signals are also found in measurements of photoion fragments (as shown in panel  \textbf{F}, reprinted from \citenum{Fehre2021}).}
    \label{fig:chirality1}
\end{figure*}

\begin{figure*}[!]
    \centering
    \includegraphics[width=0.8\textwidth]{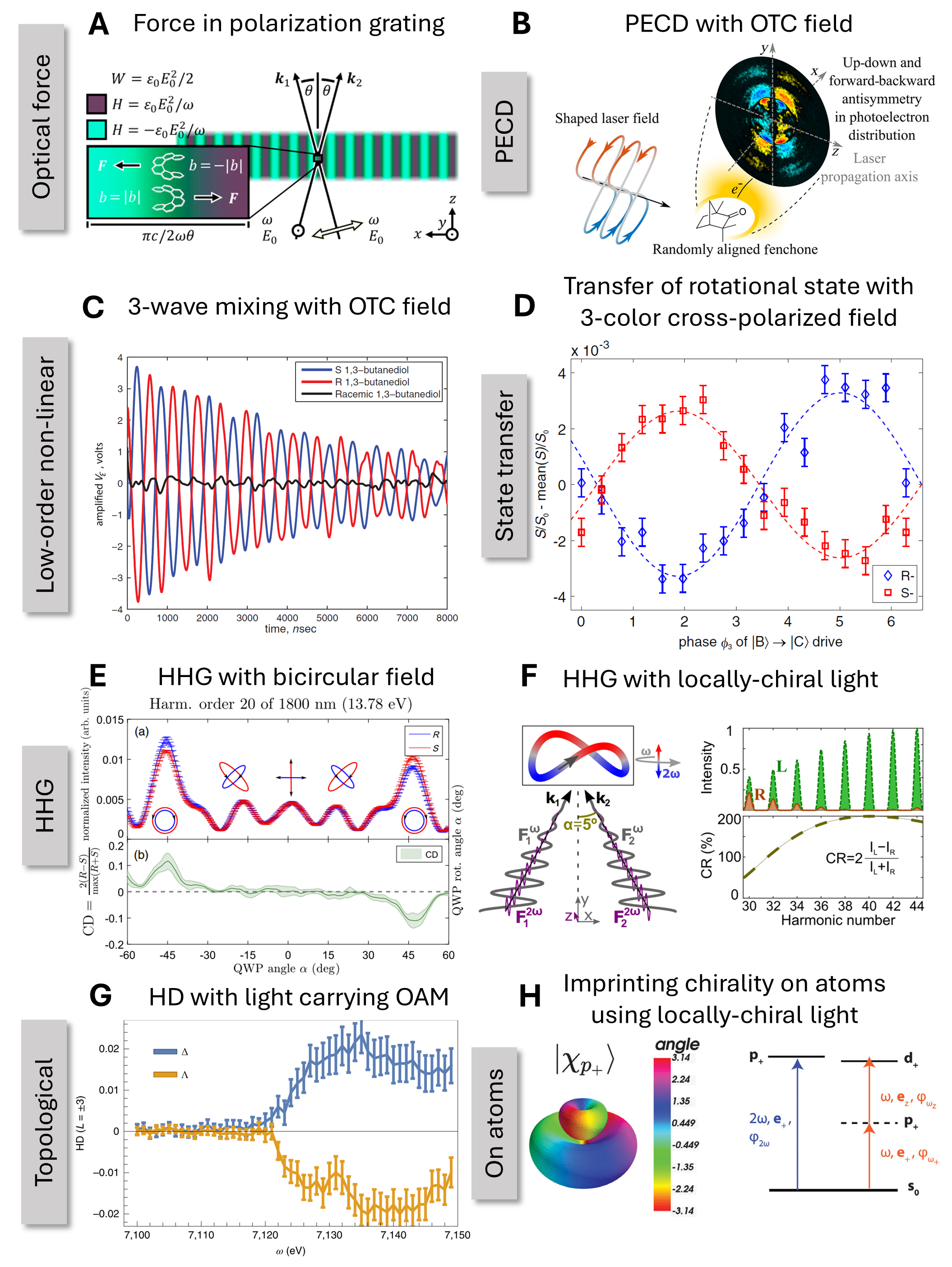}
    \caption{\textbf{Applications of tailored laser fields to investigate molecular chirality}. Panel \textbf{A} is a figure reprinted from \citenum{Cameron2014} showing a polarization grating creates an enantio-sensitive force that separates molecular enantiomers. Panel \textbf{B} is a figure reprinted from \citenum{Rozen2019}, showing an example of how PECD experiments can also benefit from using tailored fields, such as an OTC fields. OTC fields can also be used to drive low-order non-linear interactions where the 3-wave mixing signal is enantiomer-dependent, as shown in panel \textbf{C}, a figure reprinted from \citenum{Patterson2013}. Panel \textbf{D} presents a figure reprinted from \citenum{Eibenberger2017}, showing the enantio-specific transfer between molecular rotational states when using a 3-color field with crossed polarizations.  Highly non-linear interactions such as HHG can also rise to strong enantio-sensitive signals. Panel \textbf{E} shows a figure reprinted from \citenum{Baykusheva2018} where the HHG signal induced by the $\omega$--$2\omega$ field together with the corresponding CD was presented. Panel \textbf{F} presents a figure reprinted from \citenum{Ayuso2022b} that illustrates the setup for the generation of locally-chiral light (left side) and shows the enantio-sensitive HHG yield (up right) and corresponding CD (bottom right). In addition, molecular chirality can also be explored using topological light: panel \textbf{G} shows a figure reprinted from \citenum{Rouxel2022} exhibiting the HD induced by light carrying OAM. Finally, locally-chiral light also serves as a tool to imprint chirality on atoms, as shown in \textbf{H}, which is a figure adapted from \citenum{mayer2022} presenting the chiral superposition of atomic electronic states.}
    \label{fig:chirality2}
\end{figure*}

Let us now briefly review ongoing efforts in the field of chirality, and connect these to the discussion above. This section is organized as follows. We will first review the main approaches for studying chirality using standard laser fields (i. e. standard elliptically/circularly polarized light), and we will show a summary of such techniques in Figure \ref{fig:chirality1}. Secondly, we will discuss recently developed techniques that use tailored laser fields, including those discussed in the previous section, and we will show a compilation of different methods in Figure \ref{fig:chirality2}. 

Chirality is an asymmetry property of matter, i.e. it arises when matter lacks certain symmetries. Consequently, to identify whether a certain object, crystal, or molecule, is chiral, it is enough to look at the point/space group characterizing it \cite{Bishop2012}. If that group excludes symmetry elements such as inversion, mirror planes, and improper rotations, then the object inherently cannot be superimposed onto its mirror image by rotations (see Figure \ref{fig:chirality1}A). Direct ramifications of this are, for instance, optical activity such as circular dichroism (CD)\cite{Berova2013}, meaning that circularly-polarized light has a helicity-dependent absorption cross-section in the medium. The first observations of CD and related optical activity happened already in the 19th century and carried enormous weight in the sugar and pharmaceutical industry. From a practical perspective, since chirality is omnipresent, even in our body's most fundamental constituents such as DNA, its identification and analysis are paramount. Complementing chemical methodologies, light-matter interactions based on optical activity have been the workhorse for this task for over a century\cite{Berova2013}. Typically, a CD is measured in crystal or solution using extremely accurate spectrometers, allowing extraction of the medium's enantiomeric excess (EE) that is defined by the normalized ratio of left/right-handed molecules: $(L-R)/(L+R)$, where $L/R$ are the concentrations of left/right-handed molecules. 

The origin of this linear-optical CD signal has a clear intuitive symmetry connection --- if we recall that chirality by definition is the lack of mirror symmetries, we could argue that a chiral interaction (one that discriminates the handedness of the molecule) is intuitively expected only when the light beam also shares this asymmetry. Circularly-polarized light beams exhibit a polarization (of both the electric and magnetic components) that traces a screw-like shape in space-time (see Figure \ref{fig:chirality1}B), meaning that it clearly does not exhibit mirror symmetries, leading to CD in the interaction signal. However, an important point is that this lack of mirror symmetry only arises when considering the spatial structure of the beam where the screw is traced-out. At any given singular local point in space, the beam is circularly polarized and traces a two-dimensional planar circle, which is mirror-symmetric. This means that the physical origins of typical CD signals are effects beyond the electric-dipole approximation, i.e. requiring information regarding the beam's spatial evolution. Indeed, as a result, common chiral signals are of the order of $10^{-3}-10^{-6}$, and arise from an interplay of electric and magnetic dipole interactions or electric quadrupole interactions\cite{Berova2013}. This signal can in principle be enhanced by applying light with shorter wavelength (making the spatial symmetry-breaking structure appear on length scales closer to that of the molecule) \cite{Alagna1998, Zhang2017}, as shown in Figure \ref{fig:chirality1}C, adapted from \citenum{Zhang2017}. Another option consists of attempting to enhance light's intrinsic symmetry breaking through superchiral structures \cite{Tang2010,Tang2011}. However, such strategies have other limitations connected with the difficulty in their generation and measurement. Overall, this greatly limits the scope of chiral spectroscopy, making gas phase approaches difficult, requiring accurate and costly measurement apparatuses, and not allowing access to more fundamental physical phenomena such as parity-violation\cite{Erez2023,Quack2002}. Another noteworthy point is that chiral molecules, especially large bio-molecules that are relevant as novel emerging drug candidates, can exhibit chirality in multiple stereo-centers. Such intricate details are very difficult to identify with optical activity alone, further motivating research.

A different approach to enhance the CD signal consists in resorting to highly nonlinear phenomena such as HHG, where chiral-selective interactions have also been intensely studied in the last decade. Initially, high harmonics driven by elliptically-polarized lasers were measured in gas-phase chiral molecules\cite{Cireasa2015}. The advantage of HHG is that it carries attosecond information about the interaction, allowing to explore additional physics such as the timescales of photo-ionization and photo-recombination. These measurements were conducted at low ellipticity values since the HHG yield exponentially diminishes with the driving ellipticity in the monochromatic case \cite{Kanai2007,Moller2012}. Chiral signals were of the order of ~1\% and arose mostly in the form of discrimination of the ellipticity that maximizes the HHG yield (see Figure \ref{fig:chirality1}D, adapted from \citenum{Cireasa2015}). This relatively large signal was very surprising given that the symmetry-breaking origin is still in light's spatial polarization structure, and hence the signal involves magnetic dipole interactions. In fact, the signal is this high (about two orders of magnitude larger than usual CD) simply due to the extreme non-perturbative non-linearity of HHG. 

Parallel to these findings, in the past years, the emergence of the so-called "electric-dipole revolution" \cite{Ayuso2022b} --- i.e. the extraction of huge chiral signals using schemes that rely only on electric-dipole interactions---has launched enormous theoretical and experimental efforts to develop new techniques for chiral sensing, also towards studying gas phase systems and ultrafast chirality. This has been initiated with the emergence of photoelectron CD (PECD), whereby one measures the helicity-dependent angular-, momentum-, and energy-resolved photoemission spectrum. The novelty here is that the chiral signal emerges already within the electric-dipole approximation, reaching scales of up to ~30\% (orders of magnitude larger than typical CD) \cite{bowering2001,janssen2014,Laur2012}. In PECD the mirror-symmetry-broken electric field arises in light's spatial structure. In that respect, electric-dipole interactions are intuitively expected to not yield chiral signals. However, the experimental apparatus angularly resolves the photoelectron yield, permitting electric-dipole-induced interactions to arise in resolved parts of the spectrum. If one spatially integrates then the chiral signal vanishes, but when looking at the forward/backward hemispheres of the photoemission (with respect to light's propagation axis) a chiral signal arises in the form of a forward/backward asymmetry (see Figure \ref{fig:chirality1}E, reprinted from \citenum{lux2012}). 
This effect has also been extended to nonlinear optical interactions in various configurations \cite{Beaulieu2018,Svodoba2022}, and even to the attosecond regime \cite{Comby2016, beaulieu2017, Facciala2023}. It has also been more recently observed in a similar process of laser-induced electron diffraction \cite{Rajak2024}. In addition, it has been proposed that photoionization by short few-cycle linearly polarized pulses results in the creation of enantio-sensitive electron vortices \cite{Planas2022}. Moreover, a similar concept to PECD consists in Coulomb explosion imaging whereby the photoions are measured in coincidence \cite{Pitzer2013,Fehre2021}, allowing also to directly measure the absolute handedness of molecules (i.e. if they are in $R/L$ configuration, and not only the presence of dichroism), where symmetry breaking arises due to spatial structure and the detection geometry (see Figure \ref{fig:chirality1}F, reprinted from \citenum{Fehre2021}).

So, where do tailored light pulses necessarily come in? We would argue that they are pertinent to the field due to the symmetry connection --- they permit either breaking symmetries that have not been broken before, or breaking the same symmetries but in a new way. 
In the following, we shall present a compilation of techniques using tailored light to investigate chirality. One of the first proposals for bringing tailored fields to chirality consisted in the use of polarization gratings that would exert an enatio-sensitive force on chiral molecules \cite{Cameron2014, Cameron2023}, enabling the separation of the enantiomers (see Figure~\ref{fig:chirality2}A, reprinted from \citenum{Cameron2014}). Other complex polarization spatial structures have also demonstrated to result in enantio-sensitive direction of light emission \cite{ayuso2021, Rego2023_2}.

PECD experiments can also benefit from the use of tailored light. For example, it has been shown that by utilizing bi-chromatic $\omega$-2$\omega$ fields that are co-propagating and transversely-polarized (OTC fields) the PECD spectra can be delay-dependent and carry information on the light field's instantaneous chirality \cite{Demekhin2018,Rozen2019,Neufeld_OC_2018} (see Figure~\ref{fig:chirality2}B). Bi-circular $\omega$--$2\omega$ fields were also recently employed for chiral attoclocks \cite{beaulieu2021}. While these effects do not have a different source from typical PECD, the nature of mirror symmetry breaking is different than that in circularly-polarized light, allowing more information to be extracted. 

Let us now move towards all-optical techniques using tailored light. First, low-order non-linear interactions have the advantage of requiring moderate laser intensities. These include second-order optical techniques that break the symmetry in polarization space such as sum-frequency generation or other wave mixing effects \cite{Simpson2004,Fischer2005, Vogwell2023, Leibscher2019} (including the microwave regime that has been particularly successful owing to good phase matching \cite{Patterson2013,Patterson2013b,Leibscher2019}, as shown in Figure \ref{fig:chirality2}C reprinted from \citenum{Patterson2013}) or using three resonant pulses that break the mirror symmetry in polarization and time, such as in enantio-specific state transfer measurements \cite{Eibenberger2017, Juhyeon2022} (see Figure \ref{fig:chirality2}D reprinted from \citenum{Eibenberger2017}) or the enantiosensitive free-induction decay steered by a tricolor cross-polarized field \cite{khokhlova2022}.

In the highly non-linear regime, the above-mentioned chiral HHG \cite{Cireasa2015} motivated additional work using tailored light to drive HHG --- it was known at the time that combinations of circular drivings allow conditions of nonzero HHG, e.g. if bi-chromatic bi-circular pulses are employed \cite{fleischer2014}. The hope was to use such poly-chromatic polarization-tailored pulses to repeat this experiment but in conditions where much larger dichroism would be obtained \cite{Ayuso2018}. Indeed, recently dichroism of up to ~10\% was measured in bicircular HHG \cite{Baykusheva2018,Harada2018} (see Figure~\ref{fig:chirality2}E reprinted from \citenum{Baykusheva2018}), and also utilized to track chiral chemical reactions in real time \cite{Baykusheva2019}. In the next stage, perturbative and wave-mixing methods were extended to HHG, where by tailoring the driving beams with polychromatic components the nature of the polarization-induced symmetry breaking can be more finely controlled \cite{Neufeld2019_2}. Here one can also generate electric-dipole chiral signals that arise in the yield and handedness of the harmonics. Lastly, specialized beams where light's symmetries are completely broken were developed, denoted as so-called locally-chiral light \cite{ayuso2019}. The idea here is to impose strict no-mirror symmetry groups on the light field at every local point in space \cite{Neufeld2020,Neufeld2022c}. That is, the time-dependent polarization of the electric (and magnetic) field in every given point in space traces out a spatio-temporal object that cannot be superposed onto its mirror twin. This extends the typical molecular definition of chirality to electromagnetism. A direct result of this polarization-induced symmetry breaking is chiral signals in HHG (and perturbative harmonics as well) that are driven by electric-dipole interactions and arise in the yield of harmonics, rather than in their handedness (see Figure~\ref{fig:chirality2}F, reprinted from \citenum{Ayuso2022b}). This technique has been proposed in multiple geometries, each with its own advantages/disadvantages \cite{ayuso2021,Rego2023,Vogwell2023,ayuso2022,Ayuso2022b,khokhlova2022}, and allows all-optical measurements that can be optimized to very high chiral signals, above 100\%, and potentially also resolve local stereo-centers within the chiral molecule \cite{Neufeld2022b}. Locally-chiral light beams were also proposed as useful for extending PECD methodologies, whereby the geometry of the measurement apparatus is no longer essential to obtain the electric-dipole chiral signal since the light carries the dipole-induced symmetry breaking. This leads to a breaking of the forward/backward asymmetry in PECD and with it chiral signals in the angle-integrated above-threshold ionization spectra that can also be employed for optical enantio-purification \cite{Neufeld2021}.

More recently, topological light beams carrying OAM were employed to develop new types of dichroism such as helical dichroism (HD) (see Figure~\ref{fig:chirality2}G reprinted from \citenum{Rouxel2022}). Generally, this signal appears in the linear optical response as a result of the interference between the electric dipole interaction with magnetic dipole and/or electric quadrupole interactions \cite{Green2023} and is a dichroism obtained upon inverting the beam's OAM handedness (i.e. the sign of its topological charge) rather than its polarization handedness \cite{Forbes2018,Brullot2016,Rouxel2022,Ye2019}. Helical dichroism can even appear with linearly polarized beams, decoupling the need for polarization control \cite{Forbes2022}. Its origin of symmetry breaking is still in light's spatial structure, but embedded in the phase front rather than polarization evolution. Technically, CD and HD can be combined to provide additional chiral information by controlling both degrees of freedom. In principle, nonlinear responses should also yield to HD (due to the symmetry-breaking nature being the same), which, to our knowledge, has not been generally shown or predicted yet, but only analyzed in nonlinear absorption spectroscopy \cite{Begin2023}. Moreover, combining OAM beams with locally-chiral light can produce even more powerful chiral spectroscopy schemes \cite{mayer2023}. 

In addition, recent studies show how light with evermore intricate waveforms such as three-color pulses could contribute towards an enantio-selective photochemistry via enantio-sensitive population transfers \cite{ordonez2023}. Finally, it is worth mentioning that locally-chiral light can also be used to imprint chirality in atoms \cite{mayer2022}, which are naturally achiral, by creating a superposition of electronic states that results in a chiral state \cite{ordonez2019} (see Figure~\ref{fig:chirality2}H reprinted from \citenum{mayer2022}).

\section*{Conclusions and Outlook}

In this Perspective we have reviewed both the individual and joint evolution of the fields of chirality and tailored light, with emphasis on how controlling symmetries and symmetry breaking in light fields allows for novel directions in the research of chirality. Both tailored fields and chirality have been intensely studied over the previous two decades; tailored fields due to their ability to observe and control ultrafast electron dynamics, and chirality due to its prevalence in biological molecules and importance in many interdisciplinary domains such as biochemistry. Given each of their many and varied applications, and potential ability to generate advances in laser and medical technology, we expect this trend of intense study in both domains to continue for many years to come.  However, given the strong connection and potential synergy between these two fields, in the remainder of this Perspective we suggest potential future advances that could be achieved via a combined approach to the study of tailored light and chirality.

Tailored fields offer many different capabilities beyond those of standard light (i.e. just single color, Gaussian beam, elliptically-polarized), and so have recently been more heavily applied to the study of chirality as discussed in detail above. These recent advances demonstrate the promise tailored beams offer for chiral spectroscopy, especially in ultrafast timescales, and open the path to many yet unexplored possibilities. In future years, one can expect novel tailored light forms to be even more useful in chiral spectroscopy in standard and emerging systems, but also for new ideas from the field of chirality to impact the engineering of light beams, for example, attosecond pulses generated by helical HHG with or without OAM. Emerging applications of tailored fields for the study of chirality could include: (i) pump-probe setups as often employed in femtosecond spectroscopies but allowing additional symmetry breaking from structured light, (ii) polychromatic schemes allowing for low order non-linear interactions (where weaker light intensities are sufficient) with enantio-sensitive emissions as well as control over rotational states, (iii) polarization structures enabling the use of enantio-sensitive forces or emission directions, (iv) OTC and bicircular fields used to extend the capabilities of PECD or HHG measurements, (v) the use of locally-chiral light to produce huge enantio-sensitive signals in HHG, (vi) combining helical dichroism signals with circular dichroism in nonlinear optics and HHG. Even further in the future, we can envision the use of exotic forms of OAM such as flying doughnut beams and transverse OAM polarization-vortex beams to define new chiral observables, and the use of topological light bringing topologically-protected quantities and robustness to the realm of chirality.(vii) Lastly, one can imagine that the additional degrees of freedom offered by tailored light (e.g. relative color phases, intensities, vortex order, singularity type etc.), will pave the way to studying fundamental properties of chiral molecules beyond simply accessing enantiomeric excess or detecting chiral centers. In this context, by scanning these additional and non-trivial degrees of freedom one can develop multi-dimensional spectroscopies to explore electronic correlations, charge dynamics, energy transfer, ro-vibrational behaviour, and more, in chiral systems. 

All of these avenues of research will be challenging, but rewarding, for theoretical, computational, and experimental physicists and chemists alike, yielding exciting prospects for both of the fields of chirality and tailored light, and driving technological advances.

\newpage

\bibliography{sample}

\section*{Acknowledgements}
We wish to thank the organizers of the ‘Quantum Battles in Attoscience 2023’ conference for bringing the authors together and thus making this perspective possible. D. H. gratefully acknowledges support by the Ministry for Science, Higher Education and Youth, Canton Sarajevo, Bosnia and Herzegovina and the support from the COST Action CA18222, ‘Attosecond Chemistry’. K.R.H acknowledges support from the National Science Foundation under Grant No. OAC-2311928, the Texas Advanced Computing Center Frontera Pathways allocation PHY20028, and the Advanced Cyberinfrastructure Coordination Ecosystem: Services \& Support allocation MCA08X034. O.N. gratefully acknowledges the generous support of a Schmidt Science Fellowship. L. R acknowledges funding from the European Union-NextGenerationEU and the Spanish Ministry of Universities via Margarita Salas Fellowship through the University of Salamanca.




\end{document}